\documentstyle[aaspptwo]{article}
\tighten

\lefthead{Balcells and Peletier}
\righthead{Colors of bulges}
\slugcomment{Astronomical Journal, Jan. 94}

\def\SAMPLE{2}

\def\METHODS{3}
\def\DUST{3.2}
\def\PHOTchecks{4}

\def\APPHOT{2}
\def\PHOTdat{1}
\def\COLandGRAD{3}
\def\METMOD{4}

\def\LINCHECK{1}
\def\hUR{2}
\def\COLCOLloc{3}
\def\COLvsSB{4}
\def\COLvsABSM{5}
\def\DCOLhist{6}
\def\COLvsBR0{7}
\def\GRADvsT{8}
\def\GRADvsABSMb{9}
\def\GRADvsGRAD{10}
\def\COLvsRAD{11}

\def\eg{{\it e.g.}}
\def\deg{\ifmmode^\circ\else$^\circ$\fi}

\def\etal{{\it et al.~}}
\def\deg{$^{\rm o~}$}
\def\m32{M~32}
\def\n205{NGC~205}
\def\mg2{Mg$_2$}

\def\bck{\hskip-0.35em}
\def\min#1{\ifmmode  {^{\prime}}                           
            \else    {$^{\prime}$}\fi
            \ifcat,#1{\bck}\else\null\fi\ #1}
\def\deg{\ifmmode {^{\rm o}}              
         \else {$^{\rm o}$}\fi}
\def\sec{\ifmmode {^{\prime\prime}~}       
         \else {$^{\prime\prime}~$}\fi}
\def\ref#1{{\hangindent=\parindent \hangafter=1 \par \noindent #1}}

\def\apj#1, {{\it ApJ,~}{\bf#1}, }
\def\apjlett#1, {{\it ApJL,~}{\bf#1}, }
\def\apjsupp#1,{{\it ApJS,~}{\bf#1}, }
\def\aj#1, {{\it AJ,~}{\bf#1}, }
\def\astrf#1, {{\it Astrofizika,~}{\bf#1}, }
\def\aasupp#1, {{\it AAS,~}{\bf#1}, }
\def\aa#1, {{\it AA,~}{\bf#1}, }

\def\mnras#1, {{\it MNRAS,~}{\bf#1}, }
\def\mmnras#1, {{\it MemRAS,~}{\bf#1}, }
\def\annrev#1, {{\it ARAA,~}{\bf#1}, }
\def\ass#1, {{\it Astrophys.\ Space\ Sci.~}{\bf#1}, }
\def\pasp#1, {{\it PASP,~}{\bf#1}, }

\begin{document}

\title{Colors and color gradients in bulges of galaxies}

\author{Marc Balcells and Reynier F. Peletier\altaffilmark{1}}
\affil{Kapteyn Laboratorium, postbus 800, 9700 AV Groningen, Netherlands}
\authoraddr{Kapteyn Lab, postbus 800, 9700 AV Groningen, Netherlands}

\altaffiltext{1}{also: European Southern Observatory, Garching, Germany}

\begin{abstract}

We have obtained surface photometry in $U$, $B$, $R$ and $I$ for a complete
optically selected sample of 45 early-type spiral galaxies, to investigate
the colors and color gradients of spiral bulges. Color profiles in
$U-R$, $B-R$, $U-B$ and $R-I$ have been determined in wedges opening on the
semi-minor axes. Based on several criteria, like the smoothness of the color
profiles, the absence of dust lanes, and the central colors, we have
defined a subsample of 18 objects whose colors are largely unaffected by dust.
We believe such colors
are suitable for infering properties of the stellar populations of bulges.

We find that the colors of bulges are predominantly bluer than those of
ellipticals.
This result holds even when bulges are compared to ellipticals
of the same luminosity,
and indicates that bulges are younger and/or more metal-poor
than old elliptical galaxies.
Most bulges do not reach solar metallicities.

Bulges show predominantly negative color gradients (bluer outward).
For bright bulges ($M_R^{Bulge} < -20.0$),
the magnitude of the gradient increases with bulge luminosity.
For fainter bulges, gradients scatter around large negative values.
The behavior of color gradients as a function of bulge luminosity
suggests different formation mechanisms for faint and bright spheroids.
For bright bulges, the scaling of gradients with luminosity suggests
a formation process involving dissipation.
The similarity with ellipticals suggests that the formation of the disk
did not affect the stellar populations of the bulge in a major way.
For small bulges ($M_R > -20$), the existence of pronounced
color gradients suggests a different formation mechanism.
For these objects, the presence of the disk
may have severely affected the radial population distribution
in the bulge.

\end{abstract}

\section{Introduction}

The optical light of spiral galaxies originates from
two main components, a flattened
disk and a central spheroid or bulge. The relative fraction of the light in
the bulge is one of the major parameters in determining its Hubble type.
Spiral galaxies range from S0 galaxies,
with a prominent bulge, to Sd's, that almost entirely
consist of a disk.
The amount of interstellar matter, as well as the rate of
star formation activity, is greater for later type galaxies.
For a recent review of bulges, see e.g. Franx (1993).

Because bulges occupy a privileged, central position in the galaxy,
studies of bulges may provide important clues on
the formation and evolution of spiral galaxies.
Their high surface brightness makes them useful
for studies of spiral galaxies at high redshift.
The study of bulges has usually focussed on the question of whether bulges
are similar to ellipticals of the same mass.
Bulges are thought to obey the r$^{1/4}$ law (Kormendy 1982),
and seem to be rotationally supported (Kormendy \& Illingworth 1982).
However, recent studies point out observed differences.
The surface brightness profiles of bulges may be
better fitted with functions other than the $r^{1/4}$ law
(Shaw and Gilmore 1989; Kent \etal 1991 for our galaxy;
Andredakis and Sanders 1993);
strong, peanut-shape deviations from elliptical isophotes
(e.g. Shaw, Dettmar \& Barteldress 1990) are only found in bulges;
these may be associated to a strong cylindrical symmetry
in the velocity field (Shaw, Wilkinson \& Carter 1993).
The basic similarity
between bulges and ellipticals has been challenged by the conjecture
that bulges may be dynamical bars,
or may grow from instabilities in bars in the disk
(Pfenniger \& Norman 1990).

Stellar populations provide important clues
regarding the star formation history of bulges and the comparison between
bulges and ellipticals.
Initially associated with globular clusters into Population II (Baade 1944),
it is now clear that bulges are not metal-poor like globular clusters, but
metal-rich like giant ellipticals (Whitford 1978).
How similar are the stellar populations of bulges and ellipticals,
in quantitative terms?
This question has only been addressed in a systematic way
for bulges of S0 galaxies.
Ellipticals and S0 galaxies follow a color--magnitude
relation in which brighter galaxies are redder (Visvanathan \& Sandage 1977).
These authors find that the color--magnitude relation
for the integrated light of S0's and
ellipticals is the same. Larson \etal (1980) find that the spread
around the relation is larger for S0's than for ellipticals, and also
for galaxies in the field as compared to galaxies in clusters.
Bower \etal (1992) show that the spread around the color--magnitude relation
in Coma is extremely low, i.e. 0.04 mag in $U$ -- $V$.
The color--magnitude relation is often regarded as a sequence of old
galaxies, along which metallicity varies (Schweizer \etal 1990).

The situation for the colors of bulges of later type disk galaxies is
more complicated. Here, colors vary wildly from one galaxy to another.
Probably, this results from the combined effects
of dust and star formation, either in the bulge itself or
in the disk seen in projection (e.g. Frogel 1985).
There have been few attempts to measure color gradients
in spiral bulges in the past.
Wirth (1981) and Wirth \& Shaw (1983) studied colors and color gradients
in bulges using aperture photometry. They
find that bulges of late-type spirals (Sb and Sc) have color gradients
several times larger than seen in ellipticals and S0's.
Van der Kruit \& Searle (1982) measure
large color gradients in the bulge of NGC~7814, an Sa galaxy.

The previous studies, which all date from more than 10 years ago,
model the radial varation of the colors
without specific concern for the effects of dust reddening.
Taking dust reddening into account is essential
if we want the colors to contain information on the stellar populations.
This can be done today by carrying out
systematic studies on well-defined samples using CCD's.
In contrast to the extensive studies of the colors and color gradients
in ellipticals in recent years, multicolor studies of bulges
of spirals have been lacking. For this reason we have observed a sample
of early-type spirals (S0-Sbc) in $UBRI$ with the aim of deriving
stellar populations and radial gradients in bulges.
The population data allows us to investigate whether
the correlations between colors, gradients and
luminosities are similar in bulges and in ellipticals.
That would emphasize the similarities between these systems,
and would suggest that the presence, or later growth of, the disk
had little influence over the population structure of the bulge.
Conversely, the data might show that
bulge colors correlate with disk parameters.
This would suggest an interplay between the processes
of bulge and disk formation, or that bulge formation
was affected by the presence of the disk.
We want to investigate whether population gradients
in bulges of late-type spirals are indeed so much larger than in S0's,
as claimed in the studies mentioned above.
To answer any of these questions we need to verify
that we can separate the
changes in metallicity and age from the effects of reddening by dust.

We find that the colors of bulges are somewhat
bluer than the colors of ellipticals of the same luminosity.
The fact that they fall near the color--magnitude relation
of old elliptical populations suggests that many bulges are old,
and have lower metallicities of bulges than bright ellipticals.
The bluer colors of many bulges shows nevertheless that
a fraction of their light comes from young stars.
We conclude that many of the large color gradients measured in previous studies
are due to the effects of dust, and
that radial population gradients in bulges do not
significantly depend on the Hubble type of the parent galaxy.
Color gradients in bulges
correlate with bulge luminosity for a range
of absolute magnitudes.  Such a correlation may be
a signature of dissipative processes during formation.

Part of the analysis involves a complete modelling of the
disk and bulge components of the galaxy images.
In another paper (Peletier \& Balcells, in preparation)
we analyze the surface brightness profiles of bulges and disks,
as well as the color profiles in the disks.
In a third paper (Peletier \& Balcells, in preparation) the
data for the individual galaxies are presented in table
and figure formats.
The present paper focusses on the colors and color gradients
of the bulges.
We describe the sample and the observations in \S 2.
In \S 3 we present the method used to obtain colors and gradients,
we discuss the criteria used to identify dust effect on the color profiles,
and present the subsample used for population studies.
Comparison with published surface photometry is given in \S 4.
In \S 5 we present the main correlations between
colors, color gradients and various galaxy parameters,
and compare data for ellipticals to the bulges' data.
In \S 6 we discuss metallicities as derived from the bulges colors,
compare them to the metallicities inferred for the MW bulge
and for spiral disks, and discuss  previous results on color gradients
in relation to ours;  we discuss gradient estimates for
the MW bulge.
Conclusions and future avenues of research are given in \S 7.

\section{Sample selection and observations}

\subsection{The sample}

The sample was selected from the Uppsala General Catalog of Galaxies
(UGC; Nilson 1973).
An initial sample of 45 objects was compiled comprising all galaxies
with right ascension between $13^h$ and $24^h$,
declination above -2\deg,
galaxy type earlier than Sc, excluding barred galaxies,
apparent blue magnitude brighter than 14.0,
major axis diameter larger than 2 arcmin, absolute galactic latitude
larger than 20\deg\
and axis ratio in B larger than 1.56.  The latter corresponds
to inclinations above 50\deg\ for disks.
Subsequent examination of the individual images
during the run led to the exclusion of 12 objects:
NGC 5935, 5930, 6207, 6585, 6796, 7177, 7286, 7428, 7463, 7753,
and UGC 9483, 10713,
due to being barred, very dusty all the way to the center, irregular,
or belonging to interacting systems. NGC 7331
was observed, but needed a lot of extra attention
because of its large size, and will be dealt with separately.

The sample is listed in Table~\PHOTdat.
Columns 1 and 2 give the UGC and NGC numbers of the galaxies.
The galaxy type in column 3 is the type given
in the {\it Third Reference Catalogue of Bright Galaxies}
(de Vaucouleurs \etal 1991, hereafter RC3):
$T<0$ corresponds to S0, $T=1$ for
Sa, $T=2$ for Sab, $T=3$ for Sb, $T=4$ for Sbc, and $T=0$
for spirals without subclass.
Column 4 gives the disk ellipticity ($\epsilon = 1 - b/a$) derived from our
elliptical fits to the galaxy images.
Absolute magnitudes are derived from our $R$-band photometry
and the velocities listed in RC3,
assuming a uniform Hubble flow with
H$_0$ = 50 km~s$^{-1}$~Mpc$^{-1}$.
Bulge-to-disk ratios ($B/D$) and disk central surface brightness
($\mu_{0R}$ were derived from the
disk-bulge decomposition described in \S~\METHODS.
The parameter $h_{UR}$ is described in \S~\DUST.
The objects listed at the bottom of the table are those
deemed unsuitable for stellar population studies (see \S~\DUST).

\subsection{Observations}

The observations were done using the Prime Focus Camera of the 2.5 m INT
at the Observatorio del Roque de los Muchachos at La Palma between
June 18 and 25, 1990. We used
a coated GEC chip of 400 $\times$ 590 pixels, with a pixel size
of 0.549''. All galaxies were observed in $U$, $B$, $R$ and $I$
on the Cousins system. Typical exposure times per galaxy were
1200 s in $U$, 600 s in $B$, and 200 s in $R$ and $I$.
To avoid saturation
we often took shorter exposures in $R$ and $I$ as well.

The observations were taken under photometric conditions and
calibrated using standard stars from
Landolt (1983), with an average of six stars per night.
To check the linearity of the CCD, two tests were performed.
First, most standard stars were observed with a range of
integration times.  The count rate was found to be constant
independent of exposure time as long as the star did not
reach the digital saturation of the CCD.
Second, we compared the surface brightness profiles that were
obtained of the same galaxy, NGC~5866 in B, on two frames
of integration times of resp. 60 and 600 sec. The 600 sec frame is almost
saturated in the center.
The difference of the two surface brightness profiles, corrected
for exposure times, is given in Figure \LINCHECK.
The two profiles agree with each other within 1\%
over the whole intensity range, therefore
the photometry in this paper does not suffer from linearity problems.
Due to the presence of Saharan dust the atmospheric extinction was high.
{}From the residuals to the standard star solutions,
our photometric accuracy is estimated to be better than 0.05 mag in $R$
and $I$ and 0.10 mag in $B$ and $U$.

The CCD was cosmetically excellent, and even before flatfielding
almost no structure could be seen  on the chip. After flatfielding using
twilight sky flats, the sky background was flat to better than 0.2\%.
The seeing ranged between 1 and 1.5 arcsec FWHM during the run,
which means that we
often suffered from undersampling. Measured on the frames the
effective seeings are typically between 1.3 and 1.5 arcsec.

\section{Color profiles}

\subsection{Color profiles from wedges}

The usual approach to obtaining color profiles in elliptical galaxies
is to average the light along ellipses and to compare
the surface brightness profiles in the various bands.
Such approach may seem suitable for bulges, if the disk light
is adequately subtracted using e.g. a 2-D disk-bulge decomposition program.
However,
we found that, for our high-inclination sample,
dust in the disk often causes
the colors on one side of the bulge to be severely reddened.
To mask out the region of the dust lane and average azimuthally
is not reliable as it
rests on visual estimates of the width of the dust lane.

Given this difficulty, we chose to derive profiles on
wedge-shaped apertures opening on the semi-minor axes.
While we lose some signal with this method, the derived profiles
gain reliability as the side of the galaxy with the dust lane
can be ignored altogether.
The direction of the major axis was determined
by fitting ellipses to the galaxy's isophotes;
the position angle in the outer regions is well defined
since these galaxies are highly inclined
(see Florido \etal 1991).
After checking that the spatial scale in arcsec/pixel
was the same in all bands, we registered
the frames in different passbands by determining the pixel offsets
from at least one star observed in all bands.
The central luminosity peak of the galaxy
cannot be used for this, since its position with respect to the center
of mass can change because of dust absorption.
The surface brightness profiles in each pass-band were derived
by azimuthally averaging the light in wedges opening on
the semi-minor axes, with an opening angle of 45\deg.
Since the light in $R$ and $I$ is the least affected by absorption, the
galaxy center in $R$ was taken to be the vertex of the wedges.

We subtracted a sky background level determined in the outer regions
of each CCD image.
Since the galaxies were close to
edge-on, they did not fill the CCD frame,
and thus the sky background could
be determined to an accuracy better than 0.2\%.
The surface brightness profiles were then corrected
for redshift and Galactic extinction.
The redshift correction is small; following Whitford (1971),
we use $\Delta U$ = --4z, $\Delta B$ = --5z, $\Delta R$ = --z and
$\Delta I$ = --z. The galactic extinction was determined from A$_B$
in the RC3 (de Vaucouleurs \etal 1991) and the Galactic extinction law
(Rieke \& Lebofsky 1985).

The resulting color profiles for all the galaxies in the sample
are shown in Appendix A.  Filled and open symbols correspond to
each of the two semi-minor axes.
In many cases, color profiles are bumpy
on one side of the galaxy but smooth on the other.
In these cases, the bumpy side is ignored,
and the smooth side is taken to give the color profile of the bulge.
When profiles are smooth on both sides of the galaxy,
the bulge color profile is taken to be the average of the
two profiles.  Finally, for some galaxies the color profiles are bumpy
on both sides.  As discussed in \S \DUST, these galaxies are not
suitable for population studies as their colors reflect
the dust distribution more than the stellar colors.

Note that, in the color profiles,
the disk light has not been subtracted.
We mark with a vertical line the radius where,
according to our disk-bulge decomposition (see below),
50\% of the $R$--band surface brightness comes from the disk.
We believe that the errors in the colors introduced
by subtracting an exponential disk model
are larger and less controllable than those caused
by disk contamination.

We find that color profiles without bumpy structures are scale-free,
i.e., color varies linearly as a function of logarithmic radius.
To determine central colors and color gradients, these profiles
were fitted to a linear relation between an inner and an outer radius.
The inner cutoff radius is needed to avoid seeing effects on the colors,
and, as a rule of thumb,
should be more than 2 seeing FWHM (Peletier \etal 1990).
However, in many cases, the color profiles
show a constant slope all the way in to 1 seeing FWHM;
in those galaxies we took an inner radius of 1 FWHM.
In those few cases in which the color profiles are steeper in a central region
spanning several seeing FWHM, and smooth and flatter further out,
we have assumed that the central region is affected by dust,
and fit only the outer, straight part of the profiles.
Examples of this situation are
NGC 5866 and other perfectly edge-on cases.
The outer cutoff was derived by running
a bulge-disk decomposition using Kent's (1984) method (see paper 2).
We took only points for which, according to the decomposition,
more than 50\%  of the light came from the bulge.
Gradients derived from points with less than 10\% disk light
were not significantly different,
which indicates that the gradients given here are not affected
by disk contamination.
Table \COLandGRAD\ lists the central colors
and the gradients for the sample galaxies.
The errors are statistical errors of the linear fit.

\subsection{Selection of a subsample suitable for population studies}

Since we want to identify stellar population
gradients, we need criteria that indicate when a profile is
suspect of being affected by dust reddening.
For some objects, this is straightforward.
During the reduction, objects were excluded from further analysis if
we could not carry out a disk-bulge decomposition, or
their color profiles were very irregular,
or so steep toward the center that
dust reddening was the only explanation for the high gradients.
These objects are
        NGC 5362, 5587, 5675,  6361,  6368,  7311, and  7537.
Inspection of the color profiles of these galaxies
(Appendix A) makes it clear that their colors are dramatically
affected by dust.  These objects appear listed in the bottom part
of Table~\PHOTdat.
NGC 6757 is barred and thus excluded, although its color profiles are smooth.

To identify dust without resort to subjective, visual inspection,
we look for the effects of dust on the light distribution of the disk.
If the disk contains a large amount of dust,
then it is unlikely that the bulge does not suffer from it.
Dust in the disk is
indicated by a large difference between
the scale lengths of the disk in $U$ and $R$:
under the assumption that dust is well mixed with stars
in the disk, reddening causes an increase of the measured
disk scale lengths, as a bigger net amount of light
is removed from the center than the outer parts
since the center is brighter.
Therefore, dust must result in a scale length that is larger
in the $U$ band than in the $R$ band.
We thus define the parameter $h_{UR} \equiv h_U / h_R - 1$,
where $h_U$ and $h_R$ are the scale lengths of exponential fits to the disk
derived from the two-dimensional data in wavebands $U$ and $R$, respectively.
Because $h_{UR}$ is empirical, deciding a cutoff value
is always somewhat arbitrary.
In the view of the distribution
of color gradients as a function of $h_{UR}$ (Figure~\hUR),
it is probably reasonable to set the cutoff at $h_{UR} = 0.2$,
given that color gradients are more homogeneous for $h_{UR} < 0.2$,
and more heterogeneous at the high-$h_{UR}$ tail of the $h_{UR}$ distribution.
Because of the sensitivity of $h_{UR}$ to dust in the disk,
we take the objects in the high--$h_{UR}$ tail as the most
likely to be affected by dust.

Intrinsic population gradients in the disk also
contribute to a large $h_{UR}$ value. However, for S0 galaxies,
$h_{UR}$ is always below 0.12, much lower than the cutoff.
If we take this as indication of the types of population gradients
in later Hubble types, then we are probably safe in assigning
$h_{UR} > 0.2$ values to dust effects.   While a contribution from
intrinsic disk gradients cannot be excluded for the moment,
it remains the case that the sensitivity of $h_{UR}$ to dust makes it
a useful empirical discriminator for the presence of dust in the disks.

Finally, we have also excluded NGC 6504 as dusty, due to its
high color indices.  Stellar populations in bulges should in principle
not be much redder than those
in the most metal rich giant ellipticals;
therefore,  any galaxies with colors
$B-R > 1.8$ or $U-R > 2.5$ must be affected by dust,
even if their color profiles are structureless.

The final sample suitable for color studies of populations of bulges
consists of the 18 objects in the top group of Table~\PHOTdat
with $h_{UR} < 0.2$.

\section{Comparison with previous surface photometry}

We have compared
our data to the limited surface photometry
of spiral galaxies available in the literature.
Only two galaxies in our sample have previously published surface photometry.
No color profiles have been found for any object in our sample.
A comparison could be made for
NGC~5987 with Kent (1984) and for NGC~7457 with Lauer (1985).
The comparison with Kent (1984) is not straightforward, since his results
are published in the Gunn $r$ band. To convert our data to $r$,
we use the formula given by Thuan \& Gunn (1976):
$$ B ~=~ g ~+~ 0.14 ~+~ 0.63~(B-V) $$
Using the stars that define the Gunn system we find:
$$ g ~-~ r ~=~ -0.501 ~+~ 1.062~(B-V) $$
We furthermore use
$$ B-R_C ~=~ 1.613~(B-V) ~+~ 0.021\ \  (\rm ~Peletier~ \it et~ al.~
\rm 1990) $$
and obtain
$$ r ~=~ B ~+~ 0.382 ~-~ 1.049~(B-R_C) $$
To transform our data to the Johnson R-band, which is used by Lauer (1985)
we use
$$ R_J = R_C - 0.12(B-R_C) - 0.07\ \ (\rm Davis\ \it et~ al.~ \rm  1985) $$

The comparisons have been performed with elliptically averaged profiles,
and are given in Fig. \LINCHECK . The agreement with both
authors is reasonably good. The bump in the difference profile with
Kent at $\approx$ 20'' is due to a dust lane that was
interpolated over by us and not by Kent, so that he sees less light.
The zero point difference is probably due to the inaccuracy
of the color conversion.

The difference profile with Lauer (1985)
shows a slight, smooth gradient, in the sense that our photometry
is slightly brighter than his in the center, as compared to the outer
parts. A similar effect is seen in many of the comparisons between
the data of Peletier \etal (1990) (PDIDC) and
Lauer (1985) (e.g. NGC~720, NGC~4261,
NGC~4374 and NGC~4636) though not in all. These gradients could be caused
by sky subtraction errors in the outer parts, since Lauer had
a very small field of view. The fact that our seeing is
better than Lauer (1.10'' vs 1.60'') also adds to the gradient.

To check the accuracy of our colors we have compared
our colors to the aperture measurements compiled
in Longo \& de Vaucouleurs (1983, 1985) and de Vaucouleurs \& Longo (1988)
for six well observed galaxies.
We determined aparent $U$, $B$, $R$ magnitudes
in artificial circular apertures on our CCD data,
and compared them to the aperture measurements.
After excluding the differences that were larger than
0.5 mag (presumably caused by e.g. centering errors) all aperture
measurements were averaged for every galaxy and every band separately.
The differences in magnitudes between our data
and the means of the aperture photometry values are given in Table~\APPHOT.

The differences between the CCD-data and the
aperture photometry are consistent with our estimate
of the photometric errors,
namely 0.05 mag in R and I, 0.07 mag in B, and 0.10 mag in U and B.
On the average, our measurements are 0.01 mag fainter in U,
0.05 in B and 0.04 mag in R than the aperture measurements.
We conclude that no systematic color trends exist
between our data and the aperture photometry.

\section{Main relations}

\subsection{Bulge colors}

We analyze the color profiles and mean colors to address
the question of the ages and metallicities of spiral bulges.
Both metallicity and age affect the colors of a stellar
population.
We show that a reasonable interpretation
of the data is that bulges, as a class, have lower metallicities,
and a larger spread in ages,
than giant ellipticals.

We begin by studying the homogeneity of the color distributions of
spiral bulges, using color--color diagrams.
These provide distance-independent diagnostics and can reveal
remaining effects of dust reddening.
In Fig.~\COLCOLloc\ we plot $U-R$ vs $B-R$
for all galaxies which show smooth color profiles. For every galaxy the radial
color profiles are sampled at regular linear intervals.
The symbols differentiate between three kinds of galaxies:
S0's (filled squares), later-type spiral bulges (crosses),
and ellipticals (circles; from a comparison sample described below).
The reddening vector for normal Galactic dust (Rieke and Lebofsky 1985)
is also plotted.
We observe that almost all galaxies
fall on a well-defined sequence in the $U-R$ vs $B-R$ diagram.
This sequence is similar for both bulges and ellipticals,
but bulges extend to bluer colors than ellipticals.
There are three outliers: NGC~5577 and NGC~5879,
which are bluer than the rest, and NGC~6504, which is redder.
The latter seems heavily reddened in the central areas (\S \SAMPLE).
If the deviation from the main bulges' sequence is
entirely due to dust reddening, the center of NGC~6504 must have
1.3 mag of extinction in the $B$ band.
The two bluer objects are likely to contain young stars.
For the rest of the sample, the colors of the stellar populations
are quite homogeneous;
for a given $B-R$, the scatter in $U-R$ is less than 0.2
magnitudes, quite small considering that the observational uncertainties
in some galaxies reach 0.1 mag.
The scatter is somewhat larger than for bright elliptical galaxies,
which show a scatter of at most 0.1 mag in $U-R$ for a given $B-R$
(Franx \& Illingworth 1990).

We conclude that the stellar populations of bulges
must be similar to those of elliptical galaxies.
Colors are predominantly bluer in bulges,
but both bulges and ellipticals follow the same sequence
in the color--color diagram.

The color differences between bulges and  ellipticals
are more easily seen in the plot of color vs surface brightness.
By plotting against surface brightness the galaxies are scaled automatically.
And, a physical relationship might exist between
enrichment and local surface brightness in bulges,
as appears to be the case in ellipticals
(Edmunds \& Philips 1989, Carvalho \& Djorgovski 1989).
A similar effect is seen also in disks;
Vila-Costas \& Edmunds (1992) find that abundances
derived from [O/H] values in disk HII regions
correlate well with local surface brightness.
In Figure \COLvsSB\ colors profiles for the same galaxies
displayed in Figure \COLCOLloc\ are plotted against
local surface brightness.
S0 Galaxies are indicated with thick lines.
To reduce clutter in the diagrams,
the locus occupied by elliptical galaxies is shown
by plotting the average color profile for ellipticals
(in thick triangles), with error bars equal to
the dispersion of elliptical colors
for a given surface brightness. (The comparison elliptical sample
contains the 20 elliptical galaxies from Peletier \etal (1990) with the
highest-quality surface photometry; these are the galaxies that remain
after the cD galaxies, and some dusty galaxies like NGC~3665 and
NGC~3801, are removed. This sample contains NGC 720, 1052, 1600, 3377,
3379, 3605, 4278, 4374, 4387, 4406, 4472, 4478, 4551, 4636, 4649, 4697,
4889, 5813, 5831, 5845 and 7626.)
For a given surface brightness, the range in color
is significantly larger for bulges than it is for ellipticals.
The early-types (S0's), are bluer than ellipticals,
with profile slopes similar to the mean elliptical slope.
Later types are less homogeneous, both in range of colors
and in profile slopes.

The bluer colors of the profiles of early-type bulges measure
intrinsic differences between bulges and ellipticals.
It is true that the disk light has not been subtracted from our profiles,
and that S0 disks show strong indications that they are
younger and bluer than bulges (Bothun \& Gregg 1990).
However, the fact that the color differential is seen
at high as well as low surface brightness
shows that what we are seeing is not the result of disk contamination.

The color differential must result from a combination of
metallicity and age differences.
Can we sort out these two contributions?
In Fig.~\COLvsABSM\ we plot bulge colors, measured at $0.5 r_e$,
against absolute magnitude.
In the left panels, colors are plotted against bulge absolute magnitude,
useful for the comparison with ellipticals.
In the right panels, they are plotted against the absolute magnitude
of the total galaxy.  Also plotted in each panel is the
relation between luminosity and mean color within 1 $r_e$
for 'old' ellipticals and S0 derived by Schweizer and Seitzer (1992)
from data of Burstein \etal (1987) (color conversions as in \S \PHOTchecks).
Bulges fall along the
color--magnitude relation of elliptical galaxies.
The elliptical color--magnitude relation is generally understood
as a sequence of old galaxies along which metallicity varies.
Thus, the bluer colors and lower luminosities of the bulges
as compared to giant ellipticals suggests that
bulges owe their bluer colors to metallicities lower
than those of giant ellipticals.
However, there is significant scatter.  Histograms of the
deviations of bulge $U-R$ and $B-R$ colors from the main color--magnitude
relation of ellipticals are given in Figure~\DCOLhist.
The distributions of color excesses peak at negative values:
$\Delta(U-R) = -0.15 \pm 0.19$ and $\Delta(B-R) = -0.08 \pm 0.08$.
Thus, bulges are somewhat bluer than ellipticals
of the same luminosity.
(A Kolmogorov-Smirnov test rejects the hypothesis that bulge and
elliptical colors are drawn from the same population at the 99.91\%
confidence level.)
The bluer colors of bulges
as compared to ellipticals probably indicates
that, for many bulges, a fraction of the light comes from
younger stellar populations.
This is clearly the case for the two Sbc in the sample
(triangles in Fig.~\COLvsABSM), but also for older types,
such as NGC~7457, an S0.


Do bulge colors correlate with global galaxy parameters?
Figure~\COLvsABSM\ shows that the correlation between colors
and absolute magnitude improves slightly
when the total galaxy luminosity is used instead of the bulge alone.
The effect is rather small, and may be due to selection effects,
given the narrow range in total absolute magnitudes.
The better correlation might in part indicate that the total galaxy potential,
not the bulge alone,
determines the chemical enrichment of the bulge.
The lack of kinematic data for most galaxies in the sample
prevents us from studying the dependency of the colors with binding energy.
However, that global galaxy parameters may play a role in the
enrichment is suggested by the fact that bulge colors correlate with
disk central colors.
We show this relation in Figure \COLvsBR0.
Disk central colors were obtained by extrapolating the disk color profiles,
obtained from the two-dimensional surface brightness fits,
to the origin. Little contamination from bulge light is expected
in this derivation.
The correlation indicates either that bulge colors
depend on disk colors, or that both the colors in disk and bulge
are controlled by the same global parameter,
such as the galaxy luminosity or the depth of the potential well
of the galaxy.

\subsection{Color gradients}

The color profiles shown in Appendix A indicate that most bulges
get bluer towards the outer parts.  Although there is a small contribution
from a generally bluer (see paper 2) disk,
the main contribution to the gradient is intrinsic to the bulge.
Bulges, like ellipticals, show predominantly negative color gradients.

Color gradients, measured in magnitudes per decade in radius,
are plotted against galaxy type in Figure \GRADvsT.
All galaxies are plotted in this figure, irrespective
of their $h_{UR}$ values.
There is a considerable scatter, specially for later types.
If, however, the ratio of the disk scale lengths in $U$ and $R$
is used as a discriminant,
and all objects with $h_{UR} > 0.2$ (plotted in Figure \GRADvsT\
with solid symbols), which are presumably dusty, are removed,
then all gradients are approximately in the range of those of S0's.
It appears that color gradients in objects as different as
S0 and Sbc are comparable; the differences between types are
smaller than the dispersion within a given type.
The figure also suggests that
any remaining dust reddening effect in the sample with $h_{UR} < 0.2$
is also of similar intensity from S0 to Sbc,
which suggests that such effects are probably small.

As expected, color gradients are correlated  to each other.
Figure \GRADvsGRAD\  shows
$\Delta~\log(U-R)/\Delta~\log(r) ~\equiv~ \nabla(U-R)$
against $\nabla(B-R)$.
No points have been excluded in this diagram on the basis
of dust criteria.  The dusty objects include those  with
most negative gradients, as expected.
The correlations
remain well defined when dusty objects are removed.

One of the main goals of this paper is to investigate the
relation between color gradients and luminosity.
We show this relation in Figure \GRADvsABSMb.
We also show gradients for the sample of ellipticals in PDIDC
(solid circles).
When plotted against bulge absolute magnitude (left panels),
color gradients show two behaviors as a function of luminosity.
For $M_R^{Bulge} < -20.0$\ (bright bulges),
gradients become increasingly negative with bulge luminosity.
They range from 0 at $M_R^{Bulge} = -20.0$\
to $\sim -0.4$ at $M_R^{Bulge} = -22.5$.
The fainter bulges clearly deviate from this trend,
and show strongly negative gradients for magnitudes well below
$M_R^{Bulge} = -20.0$.
It is of course tempting to adscribe these strong negative
gradients to remaining dust effects, but the early type of
some of these objects argues against this interpretation.
While these bulges are small in size, all of them show smooth
color profiles in at least one of the two semi-minor axis wedges.
The errors in the absolute magnitudes arising from
the disk-bulge decomposition are unlikely to exceed
a few tenths of a magnitude for these small bulges.
Therefore, the gradients observed
suggest that different mechanisms are responsible for
the stellar population gradients of small and large bulges.

The behavior of color gradients with luminosity in bulges
shows remarkable similarities to that of gradients of ellipticals.
For the latter, gradients increase in amplitude
in the range $-20.0 > M_R^{Bulge} > -22.0$, while
fainter ellipticals ($M_R > -20.0$) can have
positive as well as negative gradients, which can be larger than the
largest gradients seen in giant ellipticals (Vader \etal 1988).
The scaling of metallicity gradients with luminosity
is predicted by models of galaxy formation
involving dissipation (e.g. Carlberg 1984).

The correlation between gradients and absolute magnitude in bulges
contrasts with the lack of trends between gradients
and galaxy type (Fig.~\GRADvsT). It seems, therefore, that
it is the luminosity of the bulge
more than the morphology of the galaxy which determines the
star formation history of these spheroidal components.

The correlation between gradients and absolute magnitude disappears when
the total galaxy magnitude is used instead of the the bulge magnitude
(compare left and right panels in Figure \GRADvsABSMb).
Similarly, no correlations exist between the bulge color gradients
and a number of other galaxy parameters, including the disk luminosity,
the bulge-to-disk ratio, the disk central surface brightness
and the disk colors.

\section{Analysis}

The colors of bulges, obtained here for the first time
free from dust reddening,
provide two basic pieces of information on the populations of bulges.
First, they
show that the colors of spiral bulges
lie along the same color--magnitude
relation as old elliptical galaxies;
while Visvanathan and Sandage (1977) include S0's in their
color--magnitude relation, our result goes beyond theirs,
since in our derivation we
control disk contamination in the derivation of the colors
and of the absolute magnitudes, which allows us to
extend the result to later type spirals than just S0.
Second, bulges
are on the average bluer than giant ellipticals.  Since
the color--magnitude relation in ellipticals
is thought to be caused by a metallicity trend,
the bluer colors suggest that, as a class,
bulges have lower metallicities than giant ellipticals.
In the following subsections we convert our colors to an abundance scale,
and contrast the result with what is known on the metallicites
of the Milky Way bulge and those of spiral disks.

\subsection{Metallicities}

We converted $B-R$ colors to an abundance scale
using the models of Peletier (1989) and the conversion from
$B-V$ to $B-R$ from PDIDC.
For single-age populations of 12 or 20 Gyr we find the metallicity -
$B-R$ conversion listed in Table~\METMOD.

According to this conversion, the colors of most of our bulges
correspond to metallicities lower than solar,
if the populations are old.
For comparison, giant ellipticals have a metallicity
of around $log(Z/Z_\odot)$ = 0.3 using the same calibration.
While the absolute calibration is uncertain, the differential measurement
with ellipticals is robust.
It implies a metalicity--magnitude relation for
spheroidal systems.
This relation has been found by other authors
(\eg Terlevich \etal 1981; Bica 1988).
Bica (1988), in his comprehensive study of the population content
of nuclei of galaxies, finds an upper metallicity of about
$\log(Z/Z_\odot) = -0.5$ for the stellar mix in low-luminosity ellipticals
and S0, consistent with the abundances we derive assuming old populations.
For other types the comparison with Bica is unclear,
as he does not provide mean abundances to compare with those
derived from color distributions, and
his spectral classes for S galaxies do not follow
Hubble type or luminosity of the bulges;
for Sa's, Bica consistently finds a higher-than-solar
maximum metallicity of the stellar population mix
($\log[Z/Z_\odot] = 0.5$), while our colors, which do not show
a trend with galaxy type, still correspond
to lower-than-solar metallicities for these galaxies
(note however that we have only one Sa in our sample).

\subsection{The bulge of the Milky Way}

Our color--metallicity conversion
argues against the belief that spiral bulges
have abundances much higher than solar.
We now analyze whether this result contrasts in any way
with what we know about abundances in
the best studied galactic bulge - the Milky Way (MW) bulge.
The abundance picture derived from the
various stellar tracers in the MW bulge is quite complex.
K and M giants
are metal rich, while RR Lyrae stars and planetary nebulae
are metal poor.
Rich (1988), working on a sample of 88 K giants
in Baade's window (BW), finds a wide range of
abundances from 0.1 to about 7 times solar,
with a mean of [Fe/H]$=0.08$, and
a related study based on a sample of
440 K giants also in BW, and using a different derivation,
concludes that the mean abundance is "approximately solar" (Sadler 1992).
Geisler \& Friel (1992), using Washington photometry,
find a mean abundance of [Fe/H]$=0.17\pm 0.55$.
These three results may not be inconsistent given the uncertainties
in the extinction in Baade's window (Tendrup 1988).
Preliminary results derived from high-resolution spectra
indicate that the mean of the abundance distribution
is almost exactly solar (Rich, priv. comm.).
Therefore, despite the very high metallicities in some of the
Bulge stars,
it is inaccurate to believe that the Bulge as a whole is
very metal rich.


Other tracers just add to the complexity of the
abundance picture of the MW bulge.
Bulge planetary nebulae studied by Ratag \etal (1982)
show oxygen abundances which peak at
[O/H]=--0.2, a value which is lower than that found in
H{\sc II} regions near the center of our Galaxy.
RR Lyrae's in the Bulge are exclusiverly metal poor,
as opposed to the Solar neighborhood, where they are metal rich
(Walker \& Terndrup 1991).

How does this picture of our Bulge compare with what we find for bulges
in other galaxies? The integrated luminosity of our Bulge has been
estimated to be L$_B$ = 2~10$^9$ L$_\odot$ (Gilmore \etal 1990),
corresponding to M$_R$ = --20.0 (assuming $B-R$ $\approx$ 1.5).
Kent \etal (1991) estimate that L$_K$ = 1.2~10$^{10}$ L$_\odot$,
corresponding to M$_R$ = --19.3 (assuming $R-K$ $\approx$ 2.5).
Assuming the Bulge to be old, and obeying the color--magnitude
relation for elliptical galaxies (Schweitzer \& Seitzer 1992), one expects
$B-R$ = 1.51 in the first case, and $B-R$ = 1.50 in the second case,
which corresponds to a metallicity [Fe/H] between --0.2 and 0,
depending on the exact age. Since Schweitzer \& Seitzer's relation
fits the external bulges reasonably well,
we do not find that the MW Bulge is in any way special.

It should be noted that the bulge of NGC~7457,
of size similar to our Bulge, is significantly bluer than would be expected
from the color-magnitude relation ($B-R$=1.39 in stead of 1.52).
Line strengths are also lower in NGC~7457 than in the Bulge (Whitford 1978).
This result stresses
that there is a non-zero scatter in the colors,
star formation history and enrichment
of galactic bulges, even at a given luminosity,
presumably caused by variations in the history of
star formation and enrichment.


\subsection{Comparison with HII regions}

This study offers a unique opportunity to compare the metallicities
derived from stellar colors
with metallicities determined from HII
regions for the region where disk and bulge overlap.
The compilation of Vila-Costas \& Edmunds (1992) contains
O/H abundances for 6 early-type spiral galaxies.
Apart from a few values in the outer parts, all metallicities
are higher than solar ($12 ~+~ log(O/H) ~=~ 8.9$).
The metallicity gradients are small:
on the average, the metallicity decreases by a factor of
10 every 18 kpc. Thus in the bulge area the gradients are
negligible. Central O/H values are around log($Z/Z_\odot$) = 0.45.
This value is much higher than that derived from the photometry
of our bulges, which is below solar (\S 6.1).
Any enrichment scenario predicts that
the gas metallicity is higher than the mean metallicity of the
underlying stellar distribution;
the latter is an average over the entire range of metallicities
of the population, while
the gas is enriched with the last generation of stars.
However, we find the difference very large.
Other processes which might contribute to such discrepancy are:
that the enrichment has continued in the disk
and not in the bulge, creating a hole in the radial distribution of
abundances; but the fact that
some HII regions studied by Vila-Costas \& Edmunds lie in the
region of the galaxies' central bulges argues against this interpretation.
Alternatively, the assumptions made when deriving metallicities from colors
may be incorrect, e.g. if the bulge stars are young and metal rich.
However, we have shown that the bulges scatter around the color--magnitude
relation for old elliptical galaxies.
Finally, we cannot exclude
inconsistencies between the methods of deriving metallicities
from colors and from O/H values in HII regions.


\subsection{Color gradients}

Wirth (1981) and Wirth and Shaw (1983) obtained colors
and color gradients for a moderate sample of spirals,
using aperture photometry.
They find that bulges show negative color gradients,
which are small in earlier-type galaxies (E, S0, Sa)
and several times larger in later types (Sb, Sc).
They interpret these gradients in terms of metallicity gradients.
All of their Sb and Sc
(NGC 891, 4261, 4244, 4565, 5746 and 5907) are edge-on spirals;
their axis ratios are larger than 4.6 (RC3). These galaxies tend to have
prominent major axis dust lanes (see e.g. van der Kruit 1991).
We have shown to what degree dust lanes affect the color profiles
of galaxies.
It is unavoidable that the colors in the central apertures
in Wirth and Shaw's measurements are severely affected by dust reddening.
This view is supported by the fact
that the central $B-V$ measured by Wirth and Shaw
is often larger than 1.1;  elliptical galaxies, which, compared to
spirals, are dust-free, are never redder than $B-V=1.03$
(Burstein \etal 1987). Such a difference between bulges and ellipticals
cannot be due to metallicity effects.

Therefore, it appears that the gradients measured by Wirth and Shaw (1983)
for late type spirals
do not trace stellar population gradients but dust reddening,
and that spiral bulges, early and late, show small population gradients
comparable to those seen in elliptical galaxies.

Some nearby galaxy bulges have been analyzed to look for population gradients.
The bulge of M~31 has been studied by Walterbos \& Kennicutt (1988).
For the least reddened side of the bulge, they find very little change
of color as a function of radius along the minor axis, consistent with
our findings. Van der Kruit \& Searle (1982), using photographic
plates, find very large color gradients between $U$, $J$ and $F$
for the edge-on Sa galaxy NGC~7814. This galaxy has been reanalyzed
by Peletier \& Knapen (1992) with optical and infrared images. They
find that the result by van der Kruit \& Searle (1982) is severely influenced
by reddening by the dust lane. The color gradients on the minor axis
outside the dust lane in all optical and optical-infrared colors
are very small, consistent with the results presented in this paper.

The situation concerning abundance gradients in the MW Bulge is unclear
at present. From shifts in the color--magnitude distribution as
a function of galactic latitude, Terndrup (1988) finds that the average
[Fe/H] in the Bulge ranges from roughly solar in BW at
$\mid b\mid = 4^\circ$ to [Fe/H] = --0.5 at $\mid b\mid = 8^\circ$.
This corresponds to
${\Delta(B-R)}\over{\Delta(log~r)}$ = --0.35, according to Table \METMOD.
If true, the metallicity gradient of our Bulge would be larger than
the gradients in all other galaxies considered here. However,
it is plausible that
this gradient is caused in part by a varying contribution from the disk,
since the disk contamination is very large.
A similar comment can be made about the results of Tyson (1991)
described by Rich (1992). From Washington photometry not much of a gradient
is seen up to z = 1 kpc, after which the average abundance starts
decreasing rapidly. This too could be due to disk contamination, along
the line of sight, or {\sl in situ}.
Terndrup \etal (1990) also see that the change in the strength of a TiO
absorption line is the largest at a field at 10$^\circ$ or 1.4 kpc,
and a similar effect is observed by Frogel \etal (1990) for the
infrared CO absorption bands.

The spatial distribution of M stars in the MW bulge (Blanco 1988)
also suggests the presence of an abundance gradient.
Classified mainly on the basis of their absorption lines,
M7, M5 and M2 stars show a vertical distribution which is much
steeper than that of the integrated $B$ and $V$ light
(from de Vaucouleurs \& Pence 1978).
This result still holds when the $K$-light profile,
which is steeper than the $B$ and $V$ profiles (Kent \etal 1991),
is used for comparison.
A metallicity gradient in the bulge could cause this steepening.
However, contamination by a disk population of super-metal rich stars
cannot be excluded here either.
As seen from H{\sc II}  regions, the disk also has metallicities
that are much larger than solar.

\subsection{Formation issues}

The fact that the colors of many bulges lie along the color--magnitude relation
for old ellipticals suggests a common formation history, i.e.
that a large fraction of bulges are predominantly old,
and that their populations
were not strongly affected by the later formation of the disk.
But the bigger scatter of bulges in the color--magnitude diagram
indicates that the populations of bulges are less homogeneous
than those of ellipticals. They probably have variations
in age and metallicity for a given luminosity.
We cannot exclude the possibility
that both young ages and higher metallicities
conspire to place the bulges onto the color--magnitude sequence
of old galaxies.
The same however is true about the colors of elliptical galaxies--
they do not allow for a unique determination of both age
and metallicity.  But since, on the basis of other arguments
(e.g. Schweizer \& Seitzer 1992), we have reason to think
that the color--magnitude diagram for ellipticals is a sequence
of old galaxies, we have some reason to extrapolate this to bulges.

We interpret the fact that bulges are somewhat bluer
than ellipticals of the same luminosity (Fig.~\DCOLhist)
as an indication that the stellar populations are not entirely old,
but contain contributions from younger populations.
Clear cases are NGC~7457 (S0), and the two Sbc in the sample,
NGC~5577 and NGC~5879.

Probably, the result which carries most information
on the formation of bulges
is the relation between color gradients and bulge absolute magnitude
(Fig.~\GRADvsABSMb).
Dissipational collapse models predict
that abundance gradients increase with luminosity.
Therefore, the color gradients of bulges with absolute magnitudes
$M_R^{Bulge} < -20.0$ are consistent with dissipational collapse
scenarios.  The similar behavior of gradients in bulges and ellipticals
of this luminosity range strengthens the view that
the two types of systems share similar formation mechanisms.

Of great interest would be to see whether color gradients in bulges
follow the trend of ellipticals in the high luminosity end,
ie. for $M_R < -22$.  In this domain, gradients in ellipticals
cease to increase with luminosity, instead they scatter around
low values ($\nabla(U-R) \sim -0.2$).
The flattening of the color gradient -- luminosity
relation at M$_R$ $<$ --22.0 is usually explained by merging,
which partially dilutes the population gradients established during
dissipational collapse (e.g. Carlberg 1984).
The bulges in our sample do not reach high enough luminosities
to properly answer this question.
The small color gradient in the bulge of NGC~4594 (M104:
$M_{R, Bulge} = -23.5, \nabla(B-R) = -0.12\pm 0.05$, Hes \& Peletier 1993)
could be a sign that such flattening
of the gradient--luminosity relation does take place for massive bulges.
The color gradient of NGC~4594 is plotted with a cross in
figure~\GRADvsABSMb.

The different behavior of color gradients with luminosity
at the faint luminosity end ($M_R^{Bulge} > -20.0$)
probably traces a different formation mechanism from
the more luminous bulges. This
could be related to bar instabilities in the disk, or to
the small bulges themselves being bar structures rather than
spheroidal mass distributions (Pfenniger and Norman 1990).
Population gradients can be created if a bar in the disk funnels gas into
the center of the galaxy, and builds up the bulge via
a starburst (Pfenniger, private communication).
Of great interest will be to have quantitative predictions
on population gradients resulting from this mechanism.

\section{Conclusions and future work}

\begin{enumerate}

\item Colors and color gradients have been measured for
a complete sample of edge-on spiral bulges.  By using appropriate
criteria, we have identified objects with colors affected by dust,
and have drawn a sample suitable for population studies.

\item Comparison with published aperture photometry shows
that our absolute photometry is accurate to
0.05 mag in $R$ and $I$, 0.07 mag in $B$, and 0.1 mag in $U$.

\item The colors of bulges lie along the color--magnitude relation
of old elliptical galaxy populations, with a slight, but significant,
offset to bluer colors.

\item The metallicities inferred from the colors correspond
to below-solar metal abundances for most of the bulges in our sample.
The mean abundance of the MW bulge is consistent with
the range of abundances derived for our sample of bulges.
There appears to be a large discrepancy between metallicities
derived from colors in the bulges and metallicities derived from [O/H] ratios
in HII regions in the disks.

\item Population gradients in bulges behave distinctly different
depending on the luminosity of the bulge.
For $M_R^{Bulge} < -20.0$, color gradients scale with luminosity,
becoming increasingly negative for brighter bulges.
For $M_R^{Bulge} > -20.0$,
gradients are strongly negative, and show no correlation with luminosity.
Color gradients in bulges do not correlate with galaxy type.

\item The scaling of gradients with luminosity supports
 models for the formation of bulges involving dissipation.

\item An important number of bulges of all luminosities show
dust and signs of ongoing star formation.

\end{enumerate}

A natural extension of this work is the measurement of line strength indices
in a number of objects in our sample for calibration of
the colors in an abundance scale.  Line indices would provide
abundance information for objects which we had to exclude due
to their colors being affected by dust reddening.

The role of the binding energy as parameter of the
enrichment could not be investigated due to the lack of
central velocity dispersions and disk rotation velocities
for most of the sample.  Central dispersions
and HI rotation curves for the entire sample
are important data sets needed to progress in the study
of the old star formation history of spiral bulges.

\section*{Acknowledgements}

We thank T. van Albada, M. Franx and S. Zepf for comments
on the manuscript.
The Isaac Newon Telescope at La Palma is operated by the RGO at the
Observatorio del Roque de los Muchachos of the Instituto de Astrofisica
de Canarias. This research has made use of the NASA/IPAC
extragalactic database (NED), which is operated by the
Jet Propulsion Laboratory, Caltech, under contract with the
National Aeronautics and Space Administration, USA.

\section*{Appendix A}

In Fig.~\COLvsRAD\ we present color profiles in 4 colors of all galaxies
that appear in Table~\PHOTdat\ (see \S 3).
Profiles for each semi-minor axis are shown with different symbols.

The first 21 galaxies are the ones that are not rejected on the basis
of their color profile, appearance etc.
For the other galaxies, for some reason a good
bulge-disk decomposition was not possible.
The tables will be published in a subsequent
paper (Peletier \& Balcells, in preparation) or are available
in electronic form from the authors.

\section*{References}

\begin{list}
{}{\itemsep 0pt \parsep 0pt \leftmargin 3em \itemindent -3em}
\item Andredakis, Y. \& Sanders, R. H. 1993, \mnras{}, submitted
\item Baade, W. 1944, \apj{100}, 137
\item Balcells, M. and Peletier, R. F. 1993,
in {\it IAU Symp. 153, Galactic Bulges},
eds. H. Habing and H. deJonghe, Dordrecht:Reidel, in press
\item Bica, E., 1988, \aa{195}, 76.
\item Blanco, V. M., 1988, \aj{95}, 1400.
\item Bothun, G.D. \& Gregg, M., 1990, \apj{350}, 73
\item Burstein, D., Davies, R. L., Dressler, A., Faber, S. M.,
Stone, R. P. S., Lynden-Bell, D., Terlevich, R. \& Wegner, G., 1987,
\apjsupp{64}, 601.
\item Bower, R. G., Lucey, J. R. \& Ellis, R. S., 1992, \mnras{254}, 589.
\item Carlberg, R. G., 1984, \apj{286}, 403.
\item Carvalho, R. R. de, \& Djorgovski, S. 1989, \apj{341}, 137
\item Davies, R. L., Sadler, E. M. \& Peletier, R. F., 1993, \mnras{262}, 650.
\item Davis, L. E., Cawson, M., Davies, R. L. \& Illingworth, G., 1985,
\aj{90}, 169.
\item de Vaucouleurs, A. \& LONGO, G., 1988, {\it Catalogue of visual
and infrared photometry of galaxies from 0.5 $\mu$m to 10 $\mu$m
(1961-1985)}, Texas Monographs in Astronomy 5, Univ. of Texas.
\item de Vaucouleurs, G. \& Pence, W., 1978, \aj{83}, 1163.
\item de Vaucouleurs, G., de Vaucouleurs, A., Corwin, H. G.,
Buta, R. J., Paturel, G. \& Fougu\'e, P., 1991, {\it Third
Reference Catalogue of Bright Galaxies}, New York, Springer.
\item Edmunds, M. G., \& Phillips, S. 1989, \mnras{241}, 9p
\item Franx, M. \& Illingworth, G., 1990, \apj{359}, L41.
\item Franx, M., 1993, in {\it IAU Symp. 153, Galactic Bulges},
eds. H. Habing and H. deJonghe, Dordrecht:Reidel, in press
\item Florido, E., Battaner, E., Prieto, M., Mediavilla, E. \&
Sanchez-Saavedra, M. L., 1991, \mnras{251}, 193.
\item Frogel, J.A., 1985, \apj{298}, 528
\item Frogel, J. A., Blanco, V. M., Terndrup, D. M. \& Whitford, A. E.,
1990, \apj{353}, 494.
\item Geisler, D. \& Friel, E. D. 1992, \aj{104}, 128
\item Gilmore, G., King, I. R. \& Van der Kruit, P. C., 1990,
{\it The Milky Way as a Galaxy}, University Science Books, Mill Valley.
\item Hes, R. \& Peletier, R. F., 1993, \aa{268}, 539.
\item Kent, S. M., 1984, \apjsupp{56}, 105.
\item Kent, S. M., Dame, and Fazio, 1991, \apj{378}, 131.
\item Kormendy, J. 1977, \apj{214}, 359
\item Kormendy, J., 1982, in {\it Morphology \& Dynamics of Galaxies}, ed.
L.~Martinet and M.~Mayor, Geneva Observatory, p. 113.
\item Kormendy, J. \& Illingworth, G. D., 1982, \apj{256}, 460.
\item Landolt, A. U. 1983, \aj{88}, 439
\item Larson, R. B., 1976, \mnras{176}, 31.
\item Larson, R. B., Tinsley, B. M. \& Caldwell, C. N., 1980, \apj{237}, 692.
\item Lauer, T. R., 1985, \apjsupp{57}, 473.
\item Longo, G. \& de Vaucouleurs, A., 1983, {\it A general
catalogue of photoelectric magnitudes and colors in the U, B, V system of
3,578 galaxies brighter than the 16$^{th}$ V-magnitude (1936-1982)},
Texas Monographs in Astronomy 3, Univ. of Texas
\item Longo, G. \& de Vaucouleurs, A., 1985, {\it Supplement to the
general catalogue of photoelectric magnitudes and colors of galaxies in
the U, B, V system}, Texas Monographs in Astronomy 3a, Univ. of Texas
\item Nilson, P. 1973, Uppsala Astron. Obs. Ann., 6 (UGC)
\item Peletier, R. F., 1989, Ph. D. Thesis, Univ. of Groningen.
\item Peletier, R.F., Davies, R.L., Illingworth, G., Davis, L. \&
Cawson, M., 1990, \aj{100}, 1091 (PDIDC)
\item Peletier, R.F. and Knapen, J.H., 1992, {\it Messenger}, 70, 57
\item Pfenniger, D. \& Norman, C. 1990, \apj{363}, 391
\item Ratag, M. A., Pottasch, S. R., Dennenfeld, M. \& Menzies, J. W.,
1992, \aa{255}, 255
\item Rich, R. M. 1988, \aj{95}, 828
\item Rich, R. M., 1992, in {\it The Stellar Populations of Galaxies},
ed. B.~Barbuy \& A.~Renzini, Reidel, Dordrecht, p. 29
\item Rieke, G. H. \& Lebofsky, M. J. 1985, \apj{288}, 618
\item Sadler, E. M., 1992, in {\it The Stellar Populations of Galaxies},
ed. B.~Barbuy \& A.~Renzini, Reidel, Dordrecht, p. 41
\item Schweizer, F., Seitzer, P., Faber, S. M., Burstein, D.,
Dalle Ore, C. M. \& Gonzalez, J., 1990, \apj{364}, L33
\item Schweizer, F. \& Seitzer, P. 1992, \aj{104}, 1039.
\item Shaw, M. A., Dettmar, R.-J. \& Barteldrees, A. 1990, \aa{240}, 36
\item Shaw, M. A. and Gilmore, G. 1989, \mnras{237}, 903
\item Shaw, M. A., Wilkinson, A. and Carter, D. 1993, \aa{268}, 511
\item Terlevich, R., Davies, R. L., Faber, S. M. \& Burstein, D., 1981,
\mnras{196}, 381.
\item Terndrup, D. M., 1988, \aj{96}, 884.
\item Terndrup, D. M., Frogel, J. A. \& Whitford, A. E., 1990, \apj{357}, 453.
\item Thuan, T. X. \& Gunn, J. E., 1976, \pasp{88}, 543.
\item Tyson, N. D., 1991, Ph. D. Thesis, Columbia University.
\item Vader, J. P., Vigroux, L., Lachi\`eze-Rey, M. \& Souviron, J.,
1988, \aa{203}, 217.
\item Vila-Costas, M.B., and Edmunds, M., 1992, \mnras{259}, 121.
\item Visvanathan, N. and Sandage, A. 1977, \apj{216}, 214
\item van der Kruit, P. C., \& Searle, L., 1982, \aa{110}, 79.
\item van der Kruit, P. C. 1990, in {\it The milky way
as a galaxy} (Mill Valley: University Science books)
\item Walker, A. R. \& Terndrup, D. M., 1991, \apj{378}, 119.
\item Walterbos, R. \& Kennicutt, R., 1988, \aa{198}, 61.
\item Whitford, A. E., 1971,  \apj{169}, 215.
\item Whitford, A. E., 1978, \apj{226}, 777.
\item Wirth, A., 1981, \aj{86}, 981.
\item Wirth, A. \& Shaw, R., 1983, \aj{88}, 171.
\end{list}

\end{document}